# Semantic Draw Engineering for Text-to-Image Creation

Yang Li[1], HuaQiang Jiang[1,2], YangKai Wu[3]

*Abstract*—Text-to-image generation is conducted through Generative Adversarial Networks (GANs) or transformer models. However, the current challenge lies in accurately generating images based on textual descriptions, especially in scenarios where the content and theme of the target image are ambiguous. In this paper, we propose a method that utilizes artificial intelligence models for thematic creativity, followed by a classification modeling of the actual painting process. The method involves converting all visual elements into quantifiable data structures before creating images. We evaluate the effectiveness of this approach in terms of semantic accuracy, image reproducibility, and computational efficiency, in comparison with existing image generation algorithms.

## I. INTRODUCTION

Algorithms for generating images from text have become an important direction in current artificial intelligence research, extending beyond academic domains [1,2,3,4] to widespread applications in engineering [5,6]. Artistic creation through natural language not only represents a novel interaction mode but also lowers the barriers to creation. This is particularly beneficial for individuals across various industries who may lack artistic knowledge but possess ideas, inspiration, and personality, thereby contributing to the creation of spiritual wealth for human society. With the rapid development of Natural Language Processing (NLP) and Computer Vision (CV) technologies, generative AI models such as Stable Diffusion, Midjourney, and DALL-E3, capable of producing high-quality images from text prompts and equally providing precise descriptions based on images, have emerged.

Based on these generative model algorithms, researchers and developers have delved deep into the art of prompting, turning this field into what is now known as "Prompt Engineering" [7,8,9]. Currently, the creative process involves users first composing descriptive prompt information (themes and content) and then adjusting model parameters (size, activity, etc.) before the model computes to generate images.

However, due to the inherent characteristics of natural language descriptions, such as semantic complexity and ambiguity, most users, especially those unfamiliar with the models, find it challenging to generate effective prompts, leading to desired outputs. Additionally, considering different models have varying parameters, a lack of engineered models exists to generate high-quality results with minimal trials and to assess the quality of prompt information. Some studies [10,11] have created "magic spells" (keywords) from annotated corpora to generate prompts, but this approach is too generalized and fails to meet personalized image and precise creative needs.

Addressing the issues raised above, this paper proposes a semantic drawing engineering model for current text-to-image models. Drawing inspiration from art sketching, the creation process is decomposed into five steps: induction of descriptive information (conceptualizing the theme), picture composition (sketch outlining), content representation (detail depiction), refined description (light and shadow processing), and description iteration (correction and perfection). This is particularly aimed at users unfamiliar with artistic knowledge, such as researchers, who wish to create illustrations for their papers through abstracts, proposing a rapid engineered image generation algorithm.
.

## II. RELATED WORK

### A. Prompt Engineering

With the rapid advancement of large language models [12] and text-to-image models [2, 5, 6], prompt engineering has become the optimal mode of interaction with large models. In this mode, users design and refine prompts to alter the computational parameters of trained models for specific applications, leveraging the current model's knowledge capabilities without the need for new training processes. This model has seen extensive application across various fields, achieving significant results in natural language understanding [13], image generation [14], and logical reasoning [15].

Current research focuses on automatic prompt generation and its refinement. For instance, AutoPrompt [16] uses a gradient search algorithm to create prompts. T. Gao and others [17] employ the generative T5 model to generate prompt templates, conducting exhaustive keyword searches. PromptIDE [18] offers interactive visualization algorithms for prompt performance evaluation. For complex tasks requiring multi-step operations, Promptchainer [19] allows

YangLi and Huaqiang Jiang are eqaual contribution to the paper. Huaqiang Jiang is corresponding author (e-mail: jhq@hznu.edu.cn). [1] works with school of information science and technology, hangzhou normal university , hangzhou, china, [2] works with engineering research center of mobile health management system , ministry of education , [3] works with Hangzhou Suosi Internet.

users to interactively construct prompt chains for corresponding subtasks, increasing the transparency and controllability of large models.

These studies primarily aim at text-to-text generative models, converting their outputs into labels and quantitatively assessing prompt information performance on given datasets.

*B. Image-to-Text Generation*

The current state of research in the image-to-text description generation field is rapidly evolving. This field focuses on converting visual images into detailed and accurate text descriptions, marking an important intersection between computer vision and natural language processing. With advancements in machine learning and deep learning technologies, significant achievements have been made in image-to-text description generation, especially in image understanding, information extraction, and natural language generation. Vinyals et al. [25] initially proposed a method combining Convolutional Neural Networks (CNN) and Recurrent Neural Networks (RNN) for generating descriptions related to image content. Anderson et al. [26] introduced a bottom-up and top-down attention mechanism to more precisely focus on important parts of images and generate descriptions. The Neural Baby Talk model [27] can identify specific objects in images and generate natural language descriptions containing these object labels. Cornia used memory networks to generate image description features, improving the relevance and accuracy of generated descriptions. Anderson and others have contributed to the understanding and interpretation of image semantics in visual environments.

These studies have provided multiple solutions to the image-to-text description generation field, including improving attention mechanisms, combining different types of neural networks, and enhancing the interpretability of models.

*C. Text-to-Image Generation*

Research in the text-to-image generation field is a current focal point, aiming to automatically generate images with high relevance and visual appeal from descriptive texts. This technology has wide applications in artistic creation, game development, film production, and more. Key research directions include: a. improving the clarity and realism of images; b. enhancing the match between generated images and text descriptions. Current mainstream algorithms employ encoder-decoder architectures, learning the contextual content of input texts to generate corresponding images. The text-to-image generation method of Generative Adversarial Networks (GAN) [20] innovatively transformed natural language descriptions into related images. Subsequently, the StackGAN model [21] significantly improved image quality and detail through a two-stage generation process. Tao and others through AttnGAN [22] introduced attention mechanisms to further enhance the accuracy and quality of text-to-image generation; similarly, BigGAN [24] conducted in-depth research to improve the quality and resolution of generated images. OpenAI's Dall-E3 [23] is capable of generating complex and detailed images from simple text descriptions.

*D. Application Models*

However, the quality of text-to-image generation largely depends on natural language prompts and subjective human judgment. It requires human participation in the loop for refinement [30]. Although there are some open-source demonstrations such as:

- Stable Diffusion https://huggingface.co/spaces/stabilityai/stable-diffusion
- Midjourney https://www.midjourney.com/
- DALL·E3 https://openai.com/product/dall-e-2

These allow users to create their artistic works with natural language input, but users need to try different wordings to obtain the desired output. The controllability of such applications is currently very low, especially if the target image content requires precise representation, it is often difficult to achieve the goal, or it can be said that a lot of work is needed but not necessarily the corresponding results.

III. SYSTEM DESIGN

The main method of existing prompt engineering involves using specific keywords to control image style, using parameters to adjust the degree of stylization and abstraction level, and customizing the resolution and aspect ratio of images. Additionally, further refinement of output effects can be achieved through image prompts, weighted impacts on images and texts, exclusion of unwanted elements, and exploration of keywords related to specific camera or lighting conditions. The existing mode belongs to the method of scientific experimentation and does not conform to the logical process of traditional artistic creation. Therefore, from the perspectives of controllability, accuracy, and precision of generated results, it poses a high threshold for users.

We have modeled the entire system's process following the logical flow of real artistic creation, dividing the logical process into the following six steps:

1. Creativity: According to the main content to be expressed, the process of artistic conception and creativity is carried out.
2. Conceptualizing Theme: Based on creative content, the picture theme is conceptualized.
3. Sketch Outlining: Choose an appropriate picture composition based on the picture theme.
4. Content Representation: Each element in the picture is depicted in detail.
5. Light and Shadow Processing: Artistic enhancement through atmosphere, light and shadow, texture, and color.
6. Correction and Perfection: Based on the picture generated in the first five steps, modify the descriptive content and iterate the generated results again.

*A. Creativity*

The creativity phase is the starting point of artistic creation, involving preliminary thinking about themes and concepts. In this stage, artists or designers explore the core messages and emotions they wish to express. To lower the threshold for users in the artistic creation process, especially for the creative process in theoretical disciplines, such as creating an illustration for a paper in mathematics or philosophy, the logical process of the creative process involves inputting the main content to be expressed, collecting inspiration, and creating mind maps. This involves exploring history, culture, technology, or nature to find inspiration and triggers for creativity.

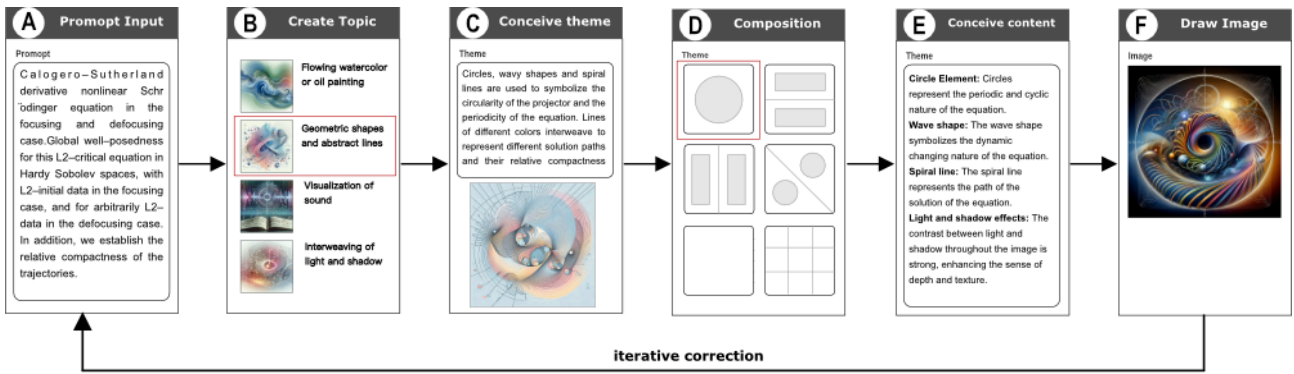

Fig. 1. Design Workflow

*B. Conceptualizing Theme*

Conceptualizing the theme is the stage of concretizing creativity, involving determining the core theme of the artwork and the mode of communication. The main logic is to refine the creativity, select themes suitable for expressing core concepts, and begin conceptualizing specific visual elements and storylines.

*C. Sketch Outlining*

Sketch outlining is the process of visualizing concepts, providing a basic framework for the final work. Sketches can typically be selected through templates or drawn by users themselves. The purpose of the sketch is to determine composition, angles, and visual layout, as well as the overall structure and balance of the work.

*D. Content Representation*

Content representation is the process of transforming concepts in sketches into detailed and specific artistic forms. The logic includes detailing each element, enhancing visual effects, and ensuring all elements are closely connected to tell a coherent story. This stage involves an in-depth understanding and application of picture elements and the interactive process.

*E. Light and Shadow Processing*

Light and shadow processing is a critical phase in the creative process, significantly impacting the emotion and atmosphere of the artwork. During this stage, the visual effect and emotional depth of each element or scene in the artwork are enhanced through adjustments in lighting, shadows, reflections, and color contrasts.

*F. Refinement and Perfection*

The final stage of refinement and perfection is a reflection and adjustment of the entire creative process to ensure the artwork achieves the desired effect. This stage involves modifying the details in the data structure generated in the previous five steps, correcting any discordant or unsatisfactory elements. This may include readjusting the composition, color balance, or the representation of details. This stage often involves peer review or client feedback to ensure the artwork meets established goals and quality standards.

Through the detailed description of these six steps, it becomes apparent that the computational process of generating images from text is not only a process of creativity and technology but also one of continuous iteration and refinement. Each stage has a decisive impact on the final effect of the artwork.

## IV. PROMPT INTERACTION MODEL

Following the ideas of the previous sections, we propose a Prompt Interaction Model which is use the SDE approach(Semantic Draw Engineering) to generate artistic images. This approach begins with the most basic information input. As revealed in Fig 2, the model's processing pipeline includes six steps: (A) Inputting the original text as the starting point of the generation process; (B) Inspiring creative thinking based on this, and selecting a specific creative style; (C) Performing hierarchical clustering analysis on the original data to form the theme of the image; (D) Selecting an appropriate compositional style from preset data templates; (E) Considering the compositional style, theme, and original data comprehensively, to perform detailed abstraction of content representation and light and shadow processing; (F) Finally generating the image and iterating the parameters to optimize the result. This process not only emphasizes the data-driven artistic creation process but also highlights the importance of the combination of creativity and technology.

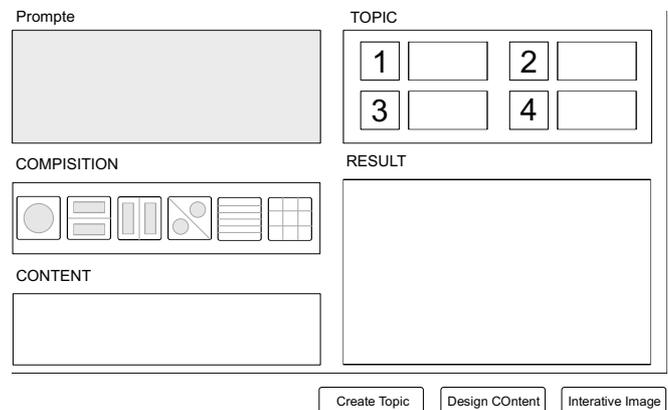

Fig. 2. Interact View

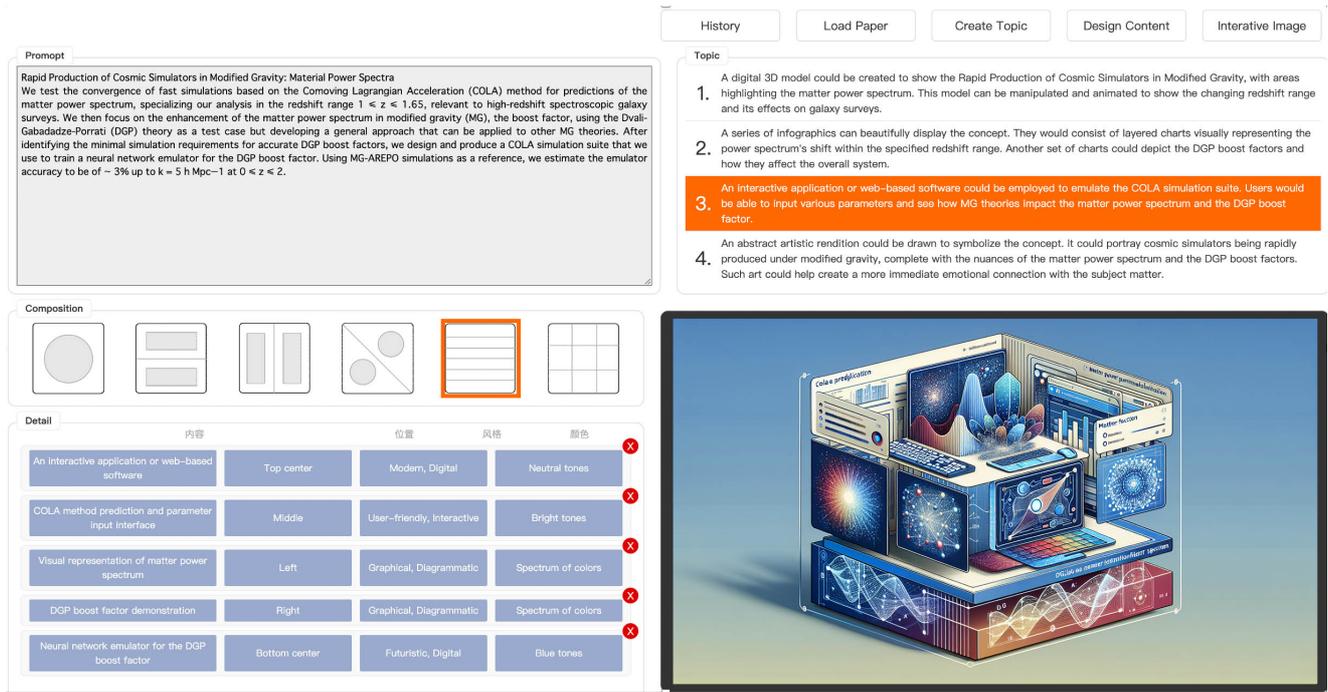

Fig. 3. Sde Approach Demo

In current training and debugging practices, we often adopt a traditional descriptive method, namely describing the image content in the form of text. This qualitative description leads to high randomness in the generated image elements and significant uncertainty in the image content. To enhance the certainty and reproducibility of image generation, this study performed in-depth quantitative data modeling on image content in process F, involving four key dimensions: element content, coordinate position, style, and color. As shown in the Fig 3, we divided the image into several areas according to the principles of composition and performed a detailed quantitative analysis of the content of each area using large models, thereby constructing the overall data structure of the image. This mode is similar to the real-world painting process, where the overall composition is first outlined with lines, then each element in the composition is refined, and finally, the details of the elements are rendered.

```
const promptTopic  = `create topic by paper's abstract`
const promptDetail = `create detail by topic and composition`

function ArtImageCreation(abstract, composition) {
  topic  = chatGpt(promptTopic)
  detail = chatGpt(promptDetail, topic, composition)
  image  = DallE3(topic, composition, detail)
  return image
}                                                    (1)
```

With above approach, to achieve a level of user satisfaction with the final product, it typically requires multiple iterations to refine the details, especially in the fifth step, step E. During this process, to expand the content of the image and enhance the saturation of the elements in the image (i.e., the richness of content), we employed the following two algorithms:

1. History Data Fusion Algorithm: This algorithm aims to merge the detail data (Detail) from historical iterations with the current Detail data. Through large-model computations, this algorithm can process and integrate multiple Detail data sets, generating a comprehensive result, as shown in Fig 4.
2. Recursive Calculation Algorithm: For the content elements of the current iteration data, this algorithm runs a recursive calculation. The purpose of this calculation is to generate a description space for a series of sub-element sets for the specified element, as shown in Fig 5.

These two algorithms not only enhance the richness of detail in the images but also strengthen the visual effect of the images while ensuring the integrity of creativity. The innovation of this method lies in its combination of the attention to detail found in traditional artistic creation and the data processing capabilities of modern computing technology, showcasing the potential of data-driven artistic creation.

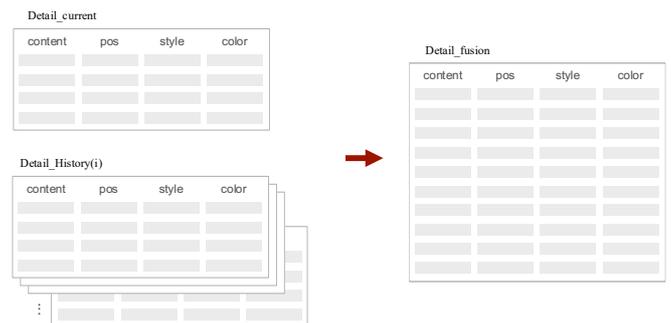

Fig.4. History Data Fusion Algorithm

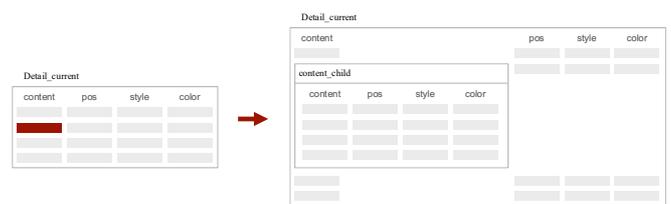

Fig.5. Recursive Calculation Algorithm

## V. USE CASES

The aforementioned study aims to provide researchers with an innovative tool for creating illustrations for papers. This tool can generate images based on the abstract or full content of a paper, aiding in explaining complex scientific definitions. The use of images is intended to simplify the understanding of complex scientific issues, especially for those researchers who are not familiar with artistic creation or lack the ability to express scientific content in a visual language.

To assess the effectiveness of this tool, we analyzed 300 scientific papers, comparing the SDE prompt algorithm with two traditional models—Midjourney and Dall-E3.

Midjourney and Dall-E3 directly use the paper's abstract as input data to generate images. Subsequently, these generated images, combined with the paper's abstract and content, were re-entered into ChatGPT to assess their performance in five aspects: conformity to the paper's theme, artistic quality of the image, understandability, computation time, and image reproducibility.

The statistical results show that the SDE prompt algorithm performs better than traditional models in multiple aspects. The specific data are as follows:

TABLE I  COMPARISON OF DIFFERENT MODEL

|  | SDE | Dall-E3 | Midjourney |
|---|---|---|---|
| Theme Conformity | 93.5% | 56.3% | 71.9% |
| Artistic Quality | 95.2% | 92.8% | 96.5% |
| Understandability | 76.7% | 53.9% | 61.3% |
| Image Reproducibility | 85.3% | 47.2% | 54.2% |
| Computation Time | 73.4s | 23.7s | 35.2s |

By comparing and analyzing the data results, the significant advantages of the SDE algorithm in assisting the generation of images for scientific papers are demonstrated. The core advantage of the SDE algorithm is its ability to accurately describe the content of papers through well-designed themes. This feature makes the algorithm outstanding in theme accuracy, and due to its use of a quantified structured data model, the reproducibility rate of the generated images is also very high, which is difficult to achieve in traditional models.

However, the SDE algorithm also has certain limitations. Specifically, due to its repeated use of large models to iterate the semantics of the paper for creative interaction, the time required to generate images is longer than traditional models. This increased time cost may be a consideration in practical applications.

Overall, despite challenges in time efficiency, the SDE algorithm still demonstrates significant advantages in improving the thematic relevance and reproducibility of illustrations for papers, especially with further improvements in large model computing power. This finding is of great significance for the practice of using images to aid in the presentation and explanation of complex concepts in the research field, providing valuable guidance for the technological development of illustrations in future scientific papers.

## VI. CONCLUSION

In this study, we convert vague semantic prompts into quantifiable data structure processes through a prompt engineering modeling approach within a text-to-image generation process. The experimental results indicate that this method, compared to traditional text-to-image prompt models, achieves higher accuracy and reproducibility in image generation, albeit with increased computational time. This research contributes to enhancing the efficacy of text prompts in the image generation process and the accuracy of semantic modeling.

ACKNOWLEDGMENT

The authors would like to thank the faculty of letters, arts and sciences of WASEDA University for the support during this research. Also they would like to give special thanks to the project of the philosophical foundation of the relationship between energy conservation, emissions reduction and socio-economic factors through AI Technology.

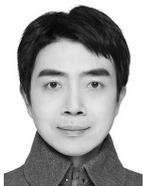
Yang Li, Ph.D. from the College of Computer Science at Zhejiang University, previously served as a postdoctoral fellow at the Oslo Institute of Water Resources in Norway. His current research interests include network services, cybersecurity, artificial intelligence, and their related technological applications, particularly in areas such as cloud computing, the Internet of Things, and big data, with a focus on energy conservation and emission reduction, assisted design, and new generative models.

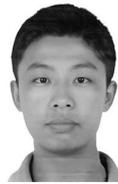
Huaqiang Jiang, holding a master's degree from Beijing University of Posts and Telecommunications. His main research areas include software engineering, artificial intelligence, and big data systems. Currently, his primary focus is on industry-academia-research collaboration, working on various application projects that integrate software and hardware in cooperation with IT companies and hospitals.

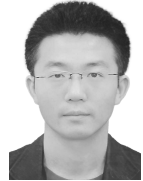
YangKai Wu, graduated from Zhejiang University majoring in Computer Science and a Master's in Computer Software from the same institution, focuses the research on automated network services, big data analysis, and autonomous learning in artificial intelligence.